\documentclass[aps,prl,twocolumn,a4paper]{revtex4}

\usepackage{amssymb, amsmath,framed}
\usepackage[colorlinks]{hyperref}
\usepackage[english]{babel}
\usepackage{graphicx}
\usepackage{graphics}
\usepackage{epstopdf}
\usepackage{epsfig}

\newcommand{\rem}[1]{}
\newcommand{\de}{{\rm d}}

\newcommand{\bx}{{\mathbf{x}}}
\newcommand{\bq}{{\mathbf{x}}}
\newcommand{\bv}{{\mathbf{v}}}
\newcommand{\bz}{{\mathbf{z}}}
\newcommand{\bbv}{{\boldsymbol{v}}}
\newcommand{\bbx}{{\boldsymbol{x}}}
\newcommand{\bbz}{{\boldsymbol{z}}}

\newcommand{\bA}{{\mathbf{A}}}

\newcommand{\bE}{{\mathbf{E}}}
\newcommand{\bB}{{\mathbf{B}}}

\newcommand{\bV}{{\boldsymbol{V}}}

\newcommand{\bX}{{\mathbf{X}}}

\newcommand{\beq}{\begin{equation}}
\newcommand{\eeq}{\end{equation}}
\newcommand{\ben}{\begin{eqnarray}}
\newcommand{\een}{\end{eqnarray}}

\def\Bbb{\mathbb}
\newcommand{\bJ}{{\mathbf{J}}}

\begin{document}

\title{Neutral Vlasov kinetic theory of magnetized plasmas}
\author{Cesare Tronci}
\email{c.tronci@surrey.ac.uk}
\affiliation{Department of Mathematics, University of Surrey, Guildford GU2 7XH, United Kingdom}
\author{Enrico Camporeale}
\email{e.camporeale@cwi.nl}
\affiliation{Center for Mathematics and Computer Science (CWI), 1098 XG Amsterdam, Netherlands}

\begin{abstract}

\smallskip

The low-frequency limit of Maxwell equations is considered in the Maxwell-Vlasov system. This limit produces a neutral Vlasov system that captures essential features of plasma dynamics, while neglecting radiation effects. Euler-Poincar\'e reduction theory is used to show that the neutral Vlasov kinetic theory possesses a variational formulation in both Lagrangian and Eulerian coordinates. By construction, the neutral Vlasov model recovers all collisionless neutral models employed in plasma simulations. Then, comparisons between the neutral Vlasov system and hybrid kinetic-fluid models are presented in the linear regime.

\end{abstract}

\maketitle



\section{Introduction}

The dynamics of magnetized plasmas is one of the most celebrated examples of multiscale systems, in which microscopic kinetic effects couple to the macroscopic scales affecting the evolution of the electromagnetic fields. This essential multiscale nature of magnetized plasmas poses well known challenges for computer simulations, which are usually required to resolve both microscopic and macroscopic scales, respectively associated to phase-space kinetics and its fluid moments. 

In the attempt to capture essential features of plasma dynamics, several  computational approaches have been proposed over the decades, based on different mathematical models. These approaches may be divided in three main categories: fully kinetic, fluid and hybrid kinetic-fluid. Each of these category may itself involve different degrees of approximation leading to different dynamic equations. 

For example, the full Maxwell-Vlasov system simulated, for instance, by means of particle-in-cell methods, may be replaced by its gyrokinetic or drift-kinetic counterparts, thereby averaging out microscopic scales involved in the particle gyromotion. On the other end, fluid treatments also possess several variants (Hall-MHD, electron MHD, extended MHD, etc.), mainly extending ideal MHD equations to incorporate different plasma features. 

All these collisionless fluid models are based on the essential hypothesis of charge neutrality, which cuts out high-frequency light wave propagation. The same hypothesis underlies the formulation of most hybrid kinetic-fluid models appearing in the literature \cite{PaBeFuTaStSu}. Many different hybrid variants are available, mainly depending on the system under consideration and on the adopted approximations. For example, in plasma fusion, hybrid MHD \cite{Cheng,Tronci2010,TrCaCaMo2014} couples the MHD bulk to a kinetic theory for energetic alpha particles. In space plasma applications, ions are typically described by the Vlasov equation, while electrons obey a fluid closure that may or may not carry inertial effects.

As mentioned above, the neutrality assumption underlying both fluid and hybrid kinetic-fluid models prevents light wave propagation. The absence of light waves in neutral models has the advantage of eliminating the need of resolving for high-frequency radiation effects, thus resulting in more efficient computational schemes. In order to eliminate radiation effects in a collisionless kinetic plasma description, one may use Darwin's model \cite{Darwin}. This a modification of the Maxwell-Vlasov system that neglects the transverse part of the displacement current, while still retaining the longitudinal  electric field. This approximation includes electrostatic and magnetostatic effects and electromagnetic induction, while eliminating light wave propagation. 
At present, the Darwin-Vlasov system is the only kinetic plasma theory that is capable of retaining essential plasma phenomena, while neglecting radiation effects without invoking charge neutrality.
However, the numerical implementation of the Darwin-Vlasov model is not straightforward, and hence  not widely used in the community (see, e.g., the discussion in \cite{Chen}). 

It is the purpose of this paper to present a new simplified kinetic theory that neglects radiation effects by assuming charge neutrality directly in the Maxwell-Vlasov system. This is done by taking the low-frequency limit $\varepsilon_0\to0$ in the Maxwell equations ($\varepsilon_0$ being the dielectric constant). This process is identical to that leading to the MHD model \cite{Freidberg} and its variants, although this is now implemented directly in the Maxwell-Vlasov system, rather than in its two fluid closure. Unlike Darwin-Vlasov, electrostatic Langmuir waves are eliminated in the new model, which yet recovers all collisionless neutral plasma models. On the other hand, similarly to the Darwin-Vlasov system, the present neutral approximation of the Maxwell-Vlasov system follows from a variational principle, which ensures mathematical and physical consistency. The proposed neutral Vlasov model reads (in standard notation):
\begin{align}\label{Vlasov}
&\frac{\partial f_s}{\partial t}+\bv\cdot\frac{\partial f_s}{\partial \bq}+\frac{q_s}{m_s}\Big(\bE+\bv\times\!\bB\Big)\cdot\frac{\partial f_s}{\partial \bv}=0
\\
&
\frac{\partial\bB}{\partial t}=-\nabla\times\bE
\label{Faraday}
\\
&\mu_0^{-1}\nabla\times\bB=\sum_s q_sn_s\bV_{\!s}
\,,\quad
\sum_s q_sn_s=0\,,
\label{Amp+Gauss}
\end{align}
where the label $s$ denotes the particle species (typically, $s=i$ and $s=e$ for ions and electrons, respectively) and where we have introduced the moment notation
$n_s=\int \!f_s\,\de^3\bv$ and $\bV_{\!s}=n_s^{-1\!}\int\!\bv\, f_s\,\de^3\bv$. The above set of equations is a closed system. This is easily shown by writing Ohm's law, as it arises from the first order moment of the $s$th kinetic equation. Notice that the special choice of $s$ is irrelevant for consistency purposes and it is only a matter of convenience. For example, one can take the first order moment of the electron kinetic equation ($s=e$) to obtain Ohm's law in the form
\begin{equation}
\bE=-\bV_e\times\bB+\frac1{q_en_e}\nabla\cdot{\Bbb{P}}_e+\frac{m_e}{q_e}\!\left(\frac{\partial \bV_{\!e}}{\partial t}+\bV_{\!e}\cdot\nabla\,\bV_{\!e}\right)
,
\label{OhmsLaw}
\end{equation}
where we have introduced the pressure tensor notation 
${\Bbb{P}}_s=m_{s\!}\int(\bv-\bV_{\!s})(\bv-\bV_{\!s})\,f_s\,\de^3\bv$ and $\bV_{\!e}$ is expressed in terms of the total current $\bJ=\mu_0^{-1}\nabla\times\bB$ by making of Amp\'ere's current balance in \eqref{Amp+Gauss}.
Equivalently, one can take the first order moment of all  kinetic equations and sum over the species.

In Section II, we present the variational formulation of the neutral Vlasov model \eqref{Vlasov}-\eqref{Amp+Gauss}. This is followed, in Section III, by a discussion of some celebrated neutral models that are naturally recovered from neutral Vlasov (by its fluid or hybrid closures). 
Finally, some examples of linearized neutral Vlasov solutions are compared to both the hybrid description and the full Maxwell-Vlasov system, in Section IV.

\section{Variational formulation}
This section presents the variational formulation of the neutral Vlasov model. This is done in two stages. First, one considers Lagrangian trajectories on phase space. Second, one applies Euler-Poincar\'e reduction theory \cite{HoTr2011,HoMaRa1998} to find the corresponding Eulerian formulation. This first part is done upon considering the Maxwell-Vlasov Lagrangian \cite{Low,Brizard,CeHoHoMa,Littlejohn,Morrison,SqQiTa} in the neutral limit $\varepsilon_0\to0$, that is
\begin{multline}\label{actionprinciple}
\hspace{-.35cm}L_{\!f_{0s}}(\boldsymbol{z}_s,\dot{\boldsymbol{z}}_s,\varphi,\dot\varphi,\bA,\dot{\bA})
\\
=\sum_s\!\int\! f_{0s}(\bz_{0s})\Big({m_s}\bbv_s(\bz_{0s})\cdot\dot{\bbx}_s(\bz_{0s})
\\
+q_s\bA\big(\bbx_s(\bz_{0s})\big)\cdot\dot{\bbx}_s(\bz_{0s})
-\frac{m_s}2|\bbv_s(\bz_{0s})|^2
\\
-q_s\varphi\big(\bbx_s(\bz_{0s})\big)\Big)\,\de^6\bz_{0s}
-\frac1{2{\color{black}\mu_0}}\int\!|\nabla\times\bA|^2\,\de^3\bq
\,.
\end{multline}
Here, the density $f_{0s}(\bz_{0s})$ is the reference (time-independent) phase space density. We have denoted the phase space labels by $\bz_{0s}=(\bx_{0s},\bv_{0s})$, while 
\begin{equation}
\boldsymbol{z}_s(\bz_{0s},t)=\big(\bbx_s(\bz_{0s},t),\,\bbv_s(\bz_{0s},t)\big)
\label{Lagrtrajects}
\end{equation} 
is the Lagrangian trajectory on phase space and the index $s$ keeps track of the particle species.
 Also, the time dependence was not made explicit in the Lagrangian functional for compactness of notation. The last integral is the magnetic field energy and involves ordinary Eulerian spatial coordinates (denoted by $\bx$). This expression of the Lagrangian comes from the general form of the phase-space Lagrangian \cite{Littlejohn,SqQiTa} for the Maxwell-Vlasov system, as it is expressed in Lagrangian coordinates. The difference between the above Lagrangian and the standard phase-space Lagrangian for Maxwell-Vlasov lies in that the above expression does not carry the electric field energy term
\[
\frac{\varepsilon_0}{2}\!\int\left|\frac{\partial \bA}{\partial t}+\nabla\varphi\right|^2\,\de^3\bq
\,,
\]
which is neglected in the neutral limit $\varepsilon_0\to0$.

The equations of motion for the Lagrangian trajectories follow from the Euler-Lagrange equations
\begin{equation}\label{EL}
\frac{\partial}{\partial t}\frac{\delta L}{\delta \dot{\bbz}_s}=\frac{\delta L}{\delta \bbz_s}
\,,\qquad
\frac{\delta L}{\delta \varphi}=0
\,,\qquad
\frac{\delta L}{\delta \bA}=0
\,.
\end{equation}
where we have used the standard notation for functional derivatives. Upon making use of delta functions, the last two equations give the Lagrangian form of the neutrality relation and Amp\`ere's current balance in \eqref{Amp+Gauss}
\begin{align*}
&\hspace{-1.4cm}\sum_s q_s\! \int\!f_{0s}(\bz_{0s})\,\delta\big(\bx-\bbx_s\big(\bz_{0s},t\big)\big)\,\de^6\bz_{0s}=0
\end{align*}
\vspace{-.67cm}
\begin{multline*}
\mu_0^{-1}\nabla\times\nabla\times\bA(\bx,t)=
\\
\sum_s q_s\! \int\!\dot{\bbx}_s(\bz_{0s},t) f_{0s}(\bz_{0s})\,\delta\big(\bx-\bbx_s\big(\bz_{0s},t\big)\big)\,\de^6\bz_{0s}
\,,
\end{multline*}
while the first Euler-Lagrange equation gives
\begin{align*}
\dot{\bbx}_s =&\ \bbv_s
\\
\dot\bbv_s =&-\frac{q_s}{m_s}\big(\nabla_{\!\bbx_s}\varphi(\bbx_s,t)+{\partial_t\bA(\bbx_s,t)}\big)
\nonumber
\\
&\qquad\qquad\qquad\qquad\qquad\quad  +\frac{q_s}{m_s}\bbv_s\times\nabla_{\!\bbx_s\!}\times\bA(\bbx_s,t)
\end{align*}
where we recall the notation \eqref{Lagrtrajects} for Lagrangian trajectories.

In order to obtain the formulation in terms of Eulerian variables, we define the Lagrange-to-Euler map for the $s$th species
\[
f_s(\bz,t)=\int \! f_{0s}(\bz_{0s})\,\delta\big(\bz-\bbz_s\big(\bz_{0s},t\big)\!\big)\,\de^6\bz_{0s}
\,,
\]
where we have denoted the Eulerian phase space coordinates by $\bz=(\bx,\bv)$. Then, we use the invariance property
\[
L_{\!f_{0s}}(\boldsymbol{z}_s,\dot{\boldsymbol{z}}_s,\varphi,\dot\varphi,\bA,\dot{\bA})
=
L_{\!f_s}(\dot{\boldsymbol{z}}_s\circ\boldsymbol{z}_s^{-1},\varphi,\dot\varphi,\bA,\dot{\bA})
\]
where 
\begin{multline}
\!\!(\dot{\boldsymbol{z}}_s\circ\boldsymbol{z}_s^{-1})(\bz)
=\big(\dot\bbx\big(\boldsymbol{z}_s^{-1}(\bz,t),t\big),\dot\bbv\big(\boldsymbol{z}_s^{-1}(\bz,t),t\big)\big)
\\
=\big(\mathbf{u}_s(\bz,t),\mathbf{a}_s(\bz,t)\big)=
\bX_s(\bz,t)
\end{multline}
is the phase space vector field generating particle trajectories. One obtains the reduced Lagrangian
\begin{multline}\label{EPactionprinciple}
\ell(\bX_s,f,\varphi,\dot\varphi,\bA,\dot{\bA})=
\\
\sum_s\!\int\! f_s(\bz,t)\Big(\big({m_s}\bv+q_s\bA(\bx,t)\big)\cdot{\mathbf{u}_s(\bz,t)}-\frac{m_s}2|\bv|^2
\\
-q_s\varphi(\bx,t)\Big)\,\de^3\bq\,\de^3\bv
-\frac1{2{\mu_0}}\int|\nabla\times\bA(\bx,t)|^2\,\de^3\bq
\,.
\end{multline}
At this point, one considers the reduced Hamilton's principle $\delta\!\int_{t_1}^{t_2}\!\ell\,\de t =0$, by using the Euler-Poincar\'e variations \cite{HoTr2011,HoMaRa1998}
\begin{align}\label{variations1}
\delta\bX_{k}&=\partial_t\mathbf{Y}_s+({\bX_{s}\cdot\nabla_{\!\bz})\mathbf{Y}_{\!s}}-({\mathbf{Y}_{\!s}}\cdot\nabla_{\!\bz})\bX_{s}
\\
\delta f_s&=-\nabla\cdot(f_s\mathbf{Y}_{\!s})
\,,
\label{variations2}
\end{align}
with $\mathbf{Y}_s$ arbitrary and vanishing at the endpoints $t_1$ and $t_2$. These variations are obtained from the definition $\bX_s=\dot{\boldsymbol{z}}_s\circ\boldsymbol{z}_s^{-1}$ and the Lagrange-to-Euler map for the particle density $f_s$; one shows that $\mathbf{Y}_{\!s}=(\delta{\boldsymbol{z}}_s)\circ\boldsymbol{z}_s^{-1}$ (see \cite{HoTr2011,HoMaRa1998,CeHoHoMa,SqQiTa}). Upon using \eqref{variations1}-\eqref{variations2} in the reduced Hamilton's principle, one finds
\begin{align}
&\bX_{k}(\bx,\bv,t)=\left(\bv\,,\,\frac{q_s}{m_s}\big(\bE+\bv\times\bB\big)\right)
\end{align}
with $\bE=-\partial_t\bA-\nabla\varphi$ and $\bB=\nabla\times\bA$,
while taking the time derivative of the Lagrange-to-Euler map yields $
{\partial_t f}+\nabla_{\!\bz}\cdot(f\bX)=0$. 
Eventually, one is left with the Vlasov equations \eqref{Vlasov},
which are accompanied by the last two Euler-Lagrange equations in \eqref{EL}, thereby returning \eqref{Amp+Gauss}. The dynamics of the vector potential $\bA$ can be recovered by finding Ohm's law, for example as in \eqref{OhmsLaw}. Then, taking the curl of the latter returns Faraday's law.

\section{Special cases: hybrid models}

As pointed out in the Introduction, the neutral Vlasov model recovers all collisionless neutral plasma models appearing in the literature over the decades. Few examples are listed below.

\medskip
\noindent
{\it 1.} Neglecting electron (mean flow) inertial effects (i.e., letting $m_e/m_i\to0$ in Ohm's Law \eqref{OhmsLaw}), yields a model that is equivalent to the kinetic-multifluid model introduced by Cheng and Johnson \cite{ChengJohnson}. In this model, Ohm's law \eqref{OhmsLaw} is written in terms of the total current $\mathbf{J}$ by ignoring  terms of the order $\mathcal{O}(m_e/m_i)$ (see equation (8) in \cite{ChengJohnson}). We remark that neglecting these terms in Ohm's law destroys the variational structure, which was recently recovered  \cite{Tronci2014}  by neglecting the electron mean flow inertia in the Lagrangian \eqref{EPactionprinciple}. This procedure leads to inertial Coriolis forces that cannot be captured by other standard methods.

\medskip\noindent
{\it 2.} Consider the case with two species, i.e. $s=i,e$. If the ion kinetic equation in \eqref{Vlasov} (with $s=i$) is replaced by its fluid closure, the neutral Vlasov system returns a hybrid reconnection model proposed by Hesse and Winske \cite{HesseWinske} to capture electron pressure anisotropies. These models are obtained by a second order moment truncation of the electron kinetic equation and have been presented over the years \cite{YiWiGaBi} in two different variants, depending on whether the electron mean flow inertia is retained or not. When these terms are neglected in Ohm's law \eqref{OhmsLaw}, then the variational structure is lost and the model can be derived by truncating the electron moment hierarchy in the kinetic-multifluid model by Cheng and Johnson \cite{ChengJohnson}.  

\medskip\noindent 
{\it 3.} If the ion kinetic features are retained and the electron kinetics in \eqref{Vlasov} (with $s=e$) is replaced by its fluid closure, the neutral Vlasov model returns a hybrid model proposed by Valentini et al. \cite{Valentini} (see equations (1)-(3) and (14) therein). It can be shown that this model also possesses a variational structure. Notice, in the computational implementation, the mass ratio value $m_e/m_i$ is usually non-physical, for numerical convenience \cite{Valentini}.

\medskip\noindent
{\it 4.} When the electron inertia is neglected in the previous case, one obtains a class of widely studied hybrid models for a massless electron fluid coupled to collisionless ion kinetics \cite{WiYiOmKaQu, matthews, lipatov}. These models have been shown to have a Hamiltonian structure in \cite{Tronci2010}, while the corresponding variational structure can be derived by neglecting terms $\sim\mathcal{O}(m_e/m_i)$ in the Lagrangian for the hybrid model in \cite{Valentini}, mentioned in the previous point. 

\medskip\noindent
{\it 5.} When both ion and electron kinetics are replaced by their corresponding fluid closure, one obtains the neutral limit of the two fluid plasma model (see e.g. \cite{PaBeFuTaStSu}). In the incompressible limit, the corresponding fluid system has been studied in \cite{Degond}.

\medskip\noindent
{\it 6.} In the previous case, neglecting electron inertia yields the celebrated Hall-MHD equations. Eventually, neglecting the Hall term leads to ideal MHD, whose hybrid versions \cite{Cheng,Tronci2010,TrCaCaMo2014} are also recovered from neutral Vlasov by considering an extra species of hot particles.

\section{Linear theory results}

We recall that the standard treatment of linear plasma waves in a homogeneous magnetized plasma 
described by the Vlasov-Maxwell system is cast in the form
${\mathbf{n}\times\mathbf{n}\times \bE} + \mathcal{D}\bE = 0$,
where the dielectric tensor $\mathcal{D}$ is defined as $\mathcal{D} = {I} + \sum_s \mathbf{\chi}_s$, $\mathbf{\chi}_s$ represents the susceptibility of the species $s$, and $\mathbf{n}$ is the index of refraction vector \cite{Stix}.
By taking the neutral limit $\varepsilon_0\to 0$, one can notice that the dielectric tensor reduces to $\mathcal{D} = \sum_s \mathbf{\chi}_s$. The form of the susceptibilities depend on the particular model one employs for each individual species. In this Section we show the dispersion relations for alfven and whistler waves, at parallel and oblique propagation, comparing the standard Vlasov-Maxwell results with the results obtained with the neutral Vlasov model \eqref{Vlasov}-\eqref{Amp+Gauss} and with a hybrid model.
As customary, we consider the background magnetic field aligned to the $z$ direction, 
and the wavevector $k$ lying in the $(x,z)$ plane. We denote by $\theta$ the angle between the wavevector and the magnetic field, 
by $\omega$ the wave real frequency, and by $\gamma$ the damping rate. For simplicity, we treat an ion-electron plasma with equal
electron and ion temperatures. The plasma beta (the ratio between thermal and magnetic energy) is equal to 0.5, 
and the ratio between ion plasma and cyclotron frequency is of the order of $7\times10^3$,  which are typical values 
for, e.g., the solar wind.
\begin{figure}
 \noindent\includegraphics[width=262 pt]{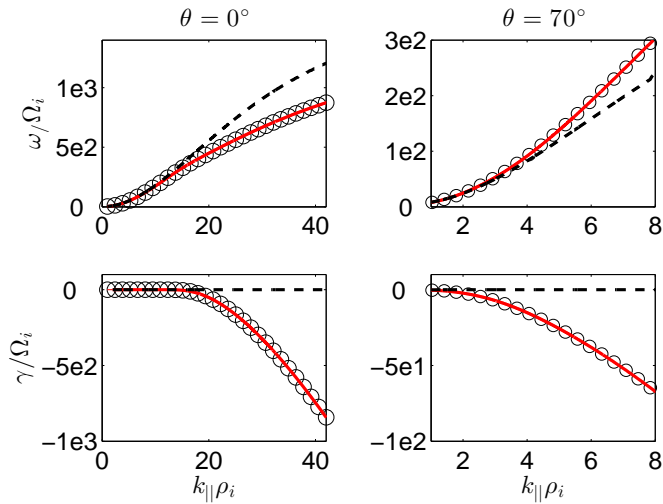}
\caption{Real frequency (top) and damping rate (bottom) for Whistler wave propagation at $\theta=0^\circ$ (left) and $\theta=70^\circ$ (right). Red line refers to neutral Vlasov, while the dashed line  and the circles are used for the hybrid model and Maxwell-Vlasov, respectively.}
\end{figure} 
In Figure 1 we show the real frequency (top panels) and the damping rate (bottom panels) as a function of the parallel wavevector $k_\parallel$ 
(normalized to the ion Larmor radius $\rho_i$), for a whistler wave. Frequencies are normalized to 
the ion cyclotron frequency $\Omega_i$. 
We have chosen an hybrid model equivalent to the one presented
in Valentini et al. \cite{Valentini}, with fluid isothermal electrons, and kinetic ions. The range of wavevectors shown emphasizes the limit of validity of hybrid models. 
Indeed, as expected, the damping due to electron kinetic is not captured in hybrid models, and already for $k_\parallel \rho_i = 4$ at oblique 
propagation there is a non-negligible mismatch with the correct Vlasov-Maxwell solution. On the other hand
 neutral Vlasov model captures the whistler dispersion relation exactly.
  \begin{figure}
 \noindent\includegraphics[width=262 pt]{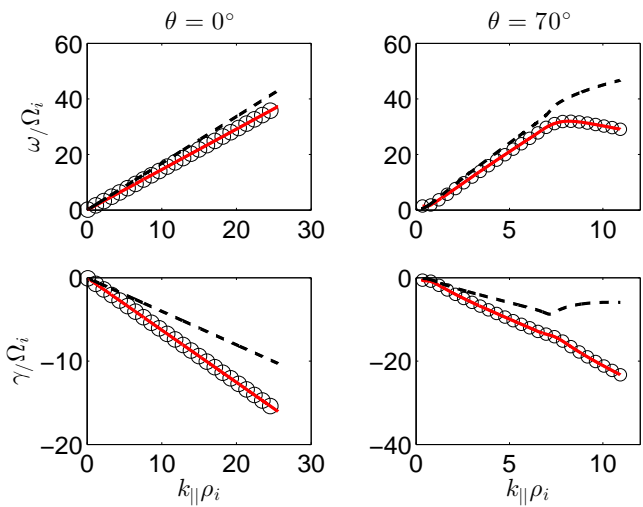}
\caption{Real frequency (top) and damping rate (bottom) for Alfv\'en wave propagation at $\theta=0^\circ$ (left) and $\theta=70^\circ$ (right). Legend is as in previous figure.
}
\end{figure} 
Figure 2 shows, in the same format as for Figure 1, the dispersion relation for Alfven waves. Once again, the neutral Vlasov 
recovers exactly the full Vlasov-Maxwell solution, both for parallel and oblique propagation. 
Although now the 
ion damping mechanisms are present in the hybrid model solution (dashed lines), one can still notice a certain mismatch. Also, an interesting
feature of branch crossing is apparent for oblique propagation (approximately at $k_\parallel\rho_i$=7), which is consistent
with the simulations presented in \cite{Vasconez}.
It is important to emphasize that although the neutral Vlasov model is computationally more expensive than the hybrid model (because
both species are treated kinetically), the mismatch in the damping rates presented in Figure 1 and 2, even at moderate $k_\parallel\rho_i$ for oblique propagation, for the hybrid model, can result in an excess of energy at small scales, which usually need to be artificially damped, for instance by using numerical filters.

\section{Conclusions}
The neutral radiationless limit of the Maxwell-Vlasov equations has been considered and the resulting neutral Vlasov system has been approached from different perspectives. The mathematical and physical consistency of the kinetic model has been supported by its variational formulations in both Lagrangian and Eulerian variables, upon using  Euler-Poincar\'e reduction in geometric mechanics \cite{HoMaRa1998}. By construction, the neutral Vlasov system recovers all collisionless neutral models appearing in the literature, some of which have been briefly discussed. The linear theory of neutral Vlasov has been compared to both its hybrid closure (with fluid electrons and kinetic ions) and the Maxwell-Vlasov system. While it has been emphasized that electrostatic Langmuir waves are lost in the neutral approximations, no mismatch was found between the fully kinetic models, for the range of wavevectors considered. 
In particular, the kinetic systems totally agree for Whistler and Alfv\`en waves at any direction of propagation. This agreement is lost between the kinetic theory and its hybrid closure, although the latter seems to capture some of the features in Alfv\`en wave propagation.
In conclusion, the neutral Vlasov model represents a promising alternative whose computational cost is in between 
hybrid and fully-kinetic models, yet exactly recovering all of the radiationless features of magnetized plasma dynamics.
For instance, it is expected that the stringent constraints due to numerical stability typical of explicit fully-kinetic codes will be relaxed, thus allowing a choice of larger timestep/ grid size. This is similar to what is achieved by the implicit moment method \cite{Lapenta2012,MaCaBuRiLa}, yet with a simpler algorithm that takes advantage of Ohm's law \eqref{OhmsLaw} to evaluate the electric field.

\smallskip

\begin{acknowledgments}
The authors are grateful to David Burgess for his keen remarks during the development of this work. Also, CT wishes to thank Pierre Degond for email correspondence on these topics. Partial support by the London Mathematical Society Grants No. 31320 \& 41371 is greatly acknowledged. 
\end{acknowledgments}

\end{document}